\documentclass[%
 aip,
 amsmath,amssymb,
 reprint,%
]{revtex4-1}

\usepackage{graphicx}
\usepackage{dcolumn}
\usepackage{bm}

\usepackage[utf8]{inputenc}
\usepackage[T1]{fontenc}
\usepackage{mathptmx}
\usepackage{etoolbox}

\newcommand{\thor}{$^{229}$Th}

\newcommand{\lisaf}{LiSrAlF$_6$}

\usepackage[dvipsnames]{xcolor}
\definecolor{ricky}{cmyk}{0, 0.7808, 0.4429, 0.1412}

\usepackage{braket}
\usepackage[colorlinks=true, allcolors=blue]{hyperref}
\usepackage{gensymb}

\usepackage{upgreek}

\usepackage{float}  
\usepackage{xcolor}

\def\bmf{BaMgF$_4$}
\def\bzf{BaZnF$_4$}
\def\gbbf{$\gamma$-Be$_2$BO$_3$F}

\usepackage[colorlinks=true, allcolors=blue]{hyperref}
\usepackage[version=3]{mhchem}

\makeatletter
\def\@email#1#2{%
 \endgroup
 \patchcmd{\titleblock@produce}
  {\frontmatter@RRAPformat}
  {\frontmatter@RRAPformat{\produce@RRAP{*#1\href{mailto:#2}{#2}}}\frontmatter@RRAPformat}
  {}{}
}%
\makeatother
\begin{document}


\title{\thor-doped nonlinear optical crystals for compact solid-state clocks}
\author{H. W. T. Morgan}
\affiliation{Department of Chemistry and Biochemistry, University of California, Los Angeles, Los Angeles, CA 90095, USA}
\affiliation{Department of Chemistry, University of Manchester, Oxford Road, Manchester M13 9PL, UK}
\author{R. Elwell}
\affiliation{Department of Physics and Astronomy, University of California, Los Angeles, CA 90095, USA}
\author{J. E. S. Terhune}
\affiliation{Department of Physics and Astronomy, University of California, Los Angeles, CA 90095, USA}
\author{H. B. Tran Tan}
\affiliation{Department of Physics, University of Nevada, Reno, Nevada 89557, USA}
\affiliation{Los Alamos National Laboratory, P.O. Box 1663, Los Alamos, New Mexico 87545, USA} 
\author{U. C. Perera}
\affiliation{Department of Physics, University of Nevada, Reno, Nevada 89557, USA}
\author{A. Derevianko}
\affiliation{Department of Physics, University of Nevada, Reno, Nevada 89557, USA}
\author{A. N. Alexandrova}
\affiliation{Department of Chemistry and Biochemistry, University of California, Los Angeles, Los Angeles, CA 90095, USA}
\author{E. R. Hudson}
\affiliation{Department of Physics and Astronomy, University of California, Los Angeles, CA 90095, USA}
\affiliation{Challenge Institute for Quantum Computation, University of California Los Angeles, Los Angeles, CA, USA}
\affiliation{Center for Quantum Science and Engineering, University of California Los Angeles, Los Angeles, CA, USA}

\date{\today}

\begin{abstract}
The recent laser excitation of the \thor{} isomeric transition in a solid-state host opens the door for a portable solid-state nuclear optical clock. 
However, at present the vacuum-ultraviolet laser systems required for clock operation are not conducive to a fieldable form factor. 
Here, we propose a possible solution to this problem by using \thor{}-doped nonlinear optical crystals, which would allow clock operation without a vacuum-ultraviolet laser system and without the need of maintaining the crystal under vacuum. 
\end{abstract}

\maketitle

The unique properties of the \thor{} nuclear isomeric state provide a laser accessible nuclear transition that can be driven even when doped into a high-bandgap solid~\cite{Hudson2008, Rellergert2010}. 
This capability is expected to allow a number of groundbreaking experiments and applications, including the construction of a robust and portable optical nuclear clock~\cite{Rellergert2010a}, exploration of nuclear superradiance~\cite{Tkalya:2010df}, tests of fundamental physics \cite{Peik_2021, Rellergert2010}, and a new probe of the solid-state environment~\cite{Gütlich2012, Elwell2024}. 

The \thor{} isomeric state was first identified in 1976 by $\gamma$-ray spectroscopy of
decaying $^{233}$U and constrained to be $< 100$~eV~\cite{Kroger1976}.
Over the ensuing decades, further experiments placed the isomer energy around 3.5~eV~\cite{helmer_reich_1994}
and several, ultimately unsuccessful, attempts~\cite{Utter1999,shaw_1999} were made to observe radiative decay of the isomeric state. 
Then in 2007-2009, an improved version of the original $\gamma$-ray spectroscopy experiment was performed and the isomer energy was revised to 7.8(5)~eV~\cite{Beck2007,Beck2009}, placing the transition squarely in the vacuum ultraviolet
(VUV) region of the spectrum and explaining why previous attempts at observing radiative decay, which were not performed in vacuum, failed. 
Since this measurement, there has
been an intense effort to directly excite the nuclear transition with the first results
reported earlier this year
\cite{Tiedau2024,Elwell2024}. 
Thus, after nearly 50 years of work, the exciting applications of the \thor{} isomeric transition are finally within reach.

Despite this important advance, challenges remain for realizing
the aforementioned goals -- particularly, realizing a portable nuclear clock system. 
The laser excitation of the isomeric state requires a rather difficult-to-achieve 148 nm laser. 
Thus far, experiments have employed either 4-wave mixing in xenon or high harmonic generation with a frequency comb to generate the requisite light and operated under high vacuum. 
While successful in exciting this transition, these rather bulky laser and vacuum systems are not particularly conducive to creating a portable nuclear clock.

Here, we detail two possible solutions~\cite{HudsonPatent2024} to this problem.
First, by doping \thor{} \emph{directly} into a nonlinear optical crystal, the requisite light can be generated and the nuclear transition interrogated \emph{without} the need for vacuum operation. 
Second, by joining a nonlinear optical crystal directly to a \thor{}-doped crystal, by e.g. optical contacting, the same capabilities can be realized. 
\begin{figure}
    \centering
    \includegraphics[width=\linewidth]{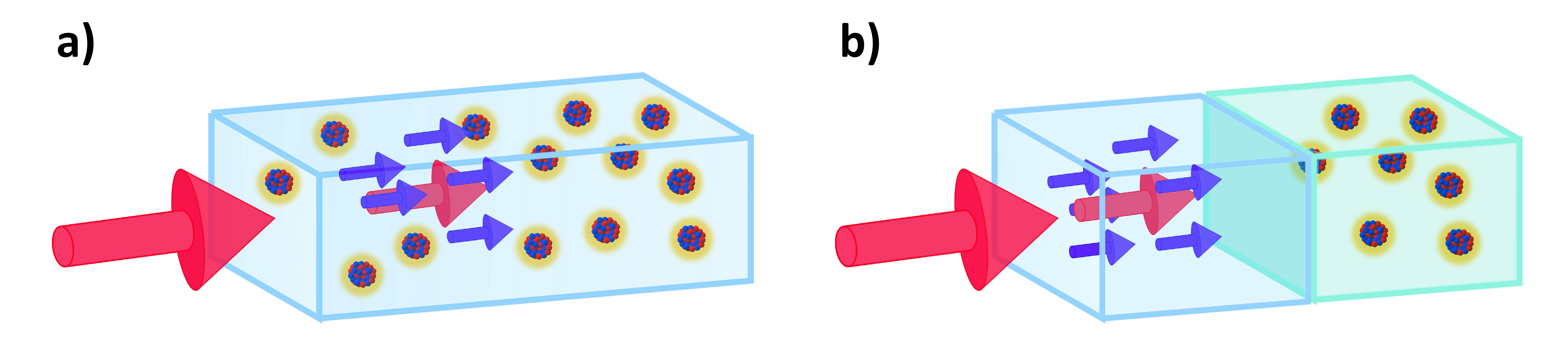}
    \caption{
    A solid-state \thor{} nuclear clock can be constructed without requiring vacuum operation by (a) doping \thor{} directly into a nonlinear optical medium or (b) joining a \thor{}-doped material directly to a nonlinear optical medium. This would allow, for example, the use of $\sim$296.8 nm light to excite the nuclear transition at $\sim$148.4 nm.
    }
    \label{fig:bmf_concept_art}
\end{figure}
While in principle straightforward, this approach clearly requires a nonlinear optical crystal capable of generating 148~nm photons, which has not yet been demonstrated.
Therefore, in what follows we first survey nonlinear optical materials that are potentially capable of generating light at 148~nm.
We then present density functional theory (DFT) calculations of the geometry and electronic structure for these crystals when doped with thorium, we also discuss their suitability for joining with other crystalline hosts of \thor{}, and conclude with a brief description of the use of these crystals in a solid-state clock. 

\section{Survey of potential nonlinear optical crystals}
The generation of higher energy photons from lower energy photons via nonlinear optical materials is an indispensable technique in modern optics. 
While ubiquitously applied throughout the visible and ultraviolet region (UV), efficient generation of photons at 148~nm using a solid-state nonlinear optical material has remained out of reach; the lowest wavelength generated by such methods is 149.8~nm~\cite{Nakazato16}. 
This deficit of NLO materials for VUV operation stems from two facts.
First, since the NLO material must be transparent to the generated light the material must have a large bandgap.
This often precludes the use of crystalline structures containing oxygen, which have so far been the most commonly employed -- e.g. BBO~\cite{Nikogosyan1991}.
Second, because the indices of refraction of material tend to increase strongly at VUV wavelengths, it is often difficult to ensure the phase matching condition can be met. 

Nonetheless, there has been a concerted, if small, effort to develop NLO materials capable of generating VUV light~\cite{Kang2022}.
Of the available materials, though fragile, \ce{KBe2BO3F2} (KBBF) has seen the most development and use.
More recently, the material $\gamma$-\ce{Be2BO3F} ($\gamma$-BBF) has been developed in hopes of solving the layering problem of KBBF~\cite{Peng2018}.
This material is expected to be capable of SHG via Type I phase matching down to a wavelength of 146~nm~\cite{Peng2018}. 

Apart from the BBF family, another promising approach is to use quasi-phase matching in periodically-poled, ferroelectric \bmf{} and \bzf{} crystals. 
\bmf{} crystals are transparent to $< 140$~nm, straightforward to pole, and appear stable under VUV illumination~\cite{Buchter2001, Herr2023}.
\bzf{} is less studied than \bmf{}, but similar properties appear possible~\cite{Villora2007}.

Thus, while other materials likely exist, the three most promising crystals at present for NLO generation of 148~nm light appear to be \gbbf{}, \bmf{}, and \bzf{}.
Therefore, in what follows we study the possible configurations for \thor{} doped into \bmf{} and \bzf{}, as well as the resulting electronic structure, to ascertain their feasibility for use in a nuclear clock where a single crystal is used both to generate the requisite VUV light and host the thorium nuclei.
We leave the case of \gbbf{} for future work. 

\section{DFT Calculations}

\subsection{\ce{BaMgF4}}

DFT is a powerful tool for studying point defects in crystalline hosts because of its high accuracy-to-cost balance, and it has been used in the development of a \thor{} crystal clock both to assess potential host crystals and to study the properties of thorium defects in thorium-free hosts~\cite{RN543,RN459,RN460,Elwell2024}.
Here we use DFT to investigate the structural and electronic properties of thorium defects in \ce{BaMgF4} to determine its potential as a clock material.
\begin{figure}
    \centering
    \includegraphics[width=0.5\textwidth]{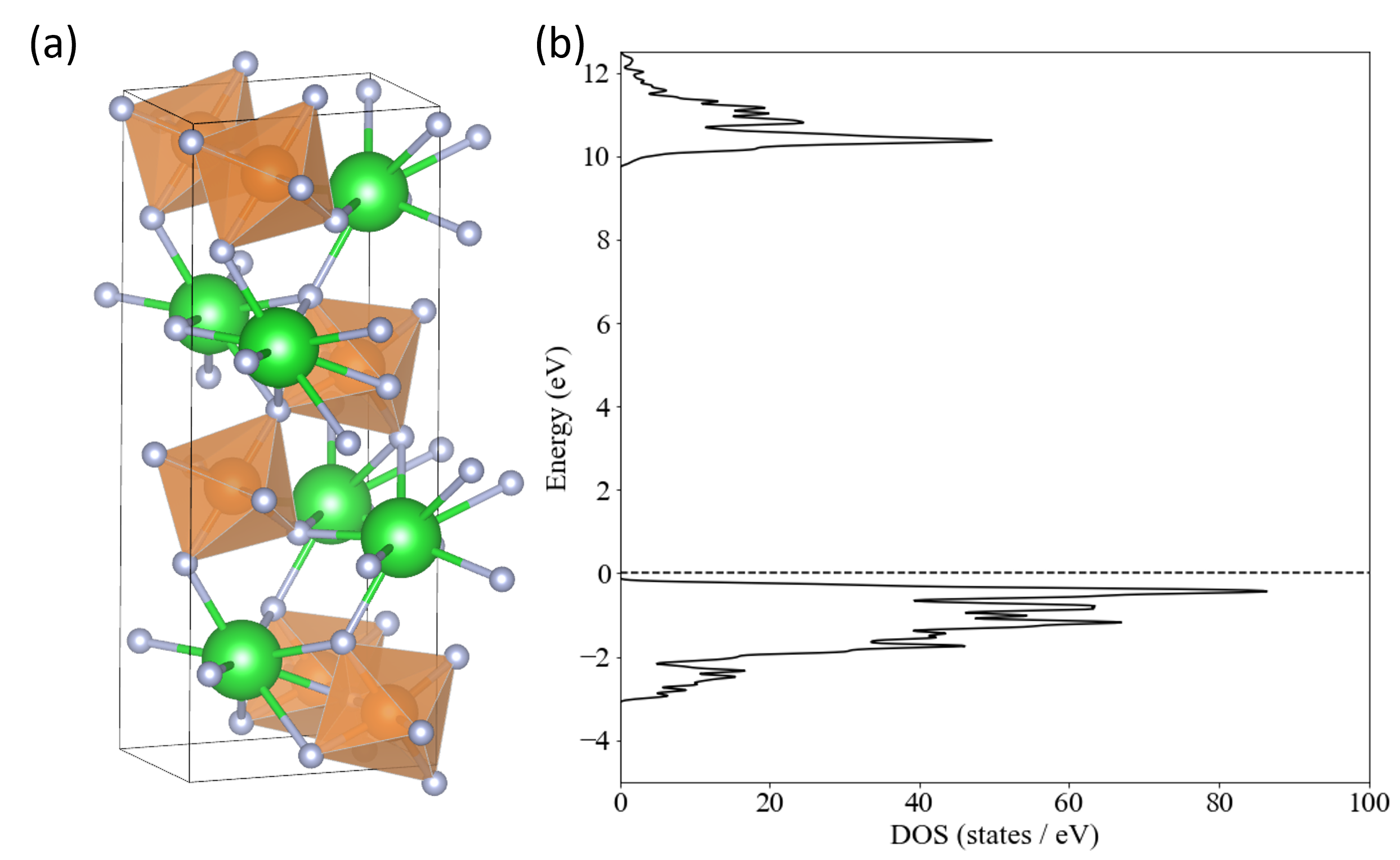}
    \caption{(a) Unit cell of \ce{BaMgF4}. Ba is shown in green, Mg in orange, and F in light grey. (b) Total density of states (DOS) of \ce{BaMgF4}.}
    \label{fig:BaMgF4 bulk}
\end{figure}
We first optimized the structure of \ce{BaMgF4} ($Cmc2_1$), shown in Figure \ref{fig:BaMgF4 bulk}(a).
The lattice parameters are overestimated by 0-2\% compared to the experimental crystal structure, as is typical for GGA functionals.\cite{RN542}
We then computed the density of states (DOS) using the MBJ functional, shown in Figure \ref{fig:BaMgF4 bulk}(b).
The computed band gap is 10.03 eV, well above the \thor{} nuclear transition energy and in good agreement with a previous absorption edge measurement of 9.92 eV.\cite{RN541}
Superior methods for computing band gaps, such as GW methods or hybrid DFT functionals, may be feasible for small unit cells of pure materials but they are not practical on defect supercells.
Thus, we compute the band gap and DOS with MBJ which is accurate and still affordable for large supercells~\cite{RN489,RN490,Elwell2024}.
These results show that our DFT model performs well on \ce{BaMgF4} and that this material is a good candidate clock material.

Having studied pure \ce{BaMgF4} we move to thorium defects.
We begin our search for defect structures with the assumptions that thorium is in the +4 oxidation state, as is typical for halides and oxides, and that \ce{Th^{4+}} replaces a \ce{Ba^{2+}} in the lattice due to the higher coordination number of the \ce{Ba^{2+}} site in \ce{BaMgF4}.
This defect, denoted \ce{Th^{..}_{Ba}}, has a 2+ charge and so must be compensated by another defect with a net 2- charge.
We have considered three possible identities for this defect: a barium vacancy, \ce{V^{$\prime\prime$}_{Ba}}; a magnesium vacancy, \ce{V^{$\prime\prime$}_{Mg}}; and a pair of fluoride interstitials, \ce{2F^{$\prime$}_{i}}.
In all cases we assume that the defect pairs are localized together, i.e. that the cation vacancies or fluoride interstitials are close to thorium.
This is expected to be the lowest-energy configuration due to electrostatic attraction between oppositely charged defects.
Defect formation energies were calculated using the binary fluorides \ce{ThF4}, \ce{BaF2}, and \ce{MgF2} as reactants and side products.
Defects were studied using a $4 \times 1 \times 3$ supercell of \ce{BaMgF4} with side lengths of approximately 15 \AA{}.
The defect formation energies are 1.30 eV for \ce{Th^{..}_{Ba} + V^{$\prime\prime$}_{Mg}}, 1.70 eV for \ce{Th^{..}_{Ba} + V^{$\prime\prime$}_{Ba}}, and 1.96 eV for \ce{Th^{..}_{Ba} + 2F^{$\prime$}_{i}}.
Optimized unit cells for all three defects are shown in the supporting information, and the formation and defect energies are summarized in Table \ref{tab:BaMgF4 defects}.
\begin{table}[htb]
\begin{tabular}{c|c|c}
Defect   & $\Delta E_{f}$ (eV) & $E_d$ (eV) \\
\hline
\ce{Th^{..}_{Ba} + V^{$\prime\prime$}_{Mg}} & 1.30     & 5.97          \\
\ce{Th^{..}_{Ba} + V^{$\prime\prime$}_{Ba}} & 1.70     & 5.73         \\
\ce{Th^{..}_{Ba} + 2F^{$\prime$}_{i}}       & 1.96     & 6.50         
\end{tabular}
\caption{Calculated formation $(\Delta E_{f})$ and defect $(E_d)$ energies of thorium defects in \ce{BaMgF4}.}
\label{tab:BaMgF4 defects}
\end{table}
Due to the large energy gap between the magnesium vacancy and barium vacancy defects, we expect \ce{Th^{..}_{Ba} + V^{$\prime\prime$}_{Mg}} to be the dominant thorium environment in Th-doped \ce{BaMgF4}.
If thermal equilibrium were established at 1138 K, the melting point of \ce{BaMgF4},\cite{BMF_mp_Pollak1984} then more than 98\% of the thorium would be in \ce{Th^{..}_{Ba} + V^{$\prime\prime$}_{Mg}} defects; this is a lower bound on the population of the defect, so it should dominate observed properties associated with thorium~\cite{RN558}.

\begin{figure}
    \centering
    \includegraphics[width=0.48\textwidth]{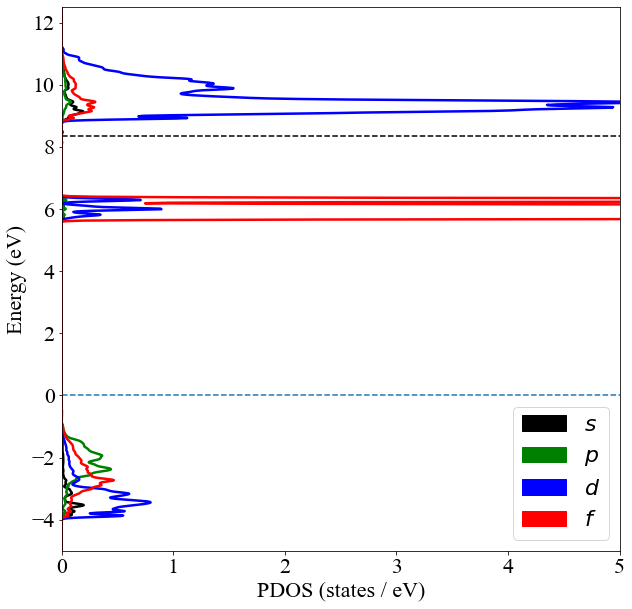}
    \caption{Thorium PDOS for the \ce{Th^{..}_{Ba} + V^{$\prime\prime$}_{Mg}} defect in \ce{Th\text{:}BaMgF4}. The dashed black line represents the energy of the \thor{} nuclear excited state relative to the top of the valence band.}
    \label{fig:VMg_Th_PDOS}
\end{figure}

The thorium-projected density of states (PDOS) for the \ce{Th^{..}_{Ba} + V^{$\prime\prime$}_{Mg}} cell is shown in Figure \ref{fig:VMg_Th_PDOS}.
The difference in energy from the Fermi energy to the first defect electronic state, $E_d$, is 5.97 eV.
However, the Th orbitals make no contribution to the uppermost part of the valence band, which suggests that the highest occupied states have no overlap with thorium.
The Th $5f$ PDOS is non-negligible below approximately -1 eV, so the lowest-energy electronic transition where both initial and final states overlap with thorium is around 7 eV.
This is likely because the occupied states at the top of the valence band are localized on fluoride ions near the magnesium vacancy that do not overlap with thorium.
Above the Th $5f$ band there is a significant energy window with zero density of states before the Th $6d$ band and bulk conduction band appear at 9 eV.

The thorium PDOS for the \ce{Th^{..}_{Ba} + V^{$\prime\prime$}_{Ba}} cell is shown in Figure S2.
For this configuration $E_d$=5.73 eV, and the Th PDOS is very similar to that in the \ce{Th^{..}_{Ba} + V^{$\prime\prime$}_{Mg}} cell.

Finally, the thorium PDOS for the \ce{Th^{..}_{Ba} + 2F^{$\prime$}_{i}} defect is shown in Figure S3.
It is broadly similar to the PDOSs of the other defects, with $E_d$=6.50 eV.
Unlike the cation vacancy cells, Th $5f$ character is present near the top of the valence band.

The PDOS of all three of these configurations are qualitatively similar to that of the lowest energy defect for \thor{}:\lisaf{} (\ce{Th^{..}_{Sr} + 2F_i^$\prime$})~\cite{Elwell2024} and \thor{}:\ce{CaF2} (\ce{Th^{..}_{Ca} + 2F_i^$\prime$})~\cite{NickersonThesis2019}. 
Given that a narrow nuclear resonance has been seen in both of these materials, \thor{}:\bmf{} appears to be a promising candidate for a compact solid-state clock. 
However, given the difficulty in growing \thor{}-doped materials, verification of this by successful observation of long-lived nuclear fluorescence following Ref.~\cite{Pineda2024} would be an important next step. 

\subsection{\ce{BaZnF4}}

\begin{figure}
    \centering
    \includegraphics[width=0.5\textwidth]{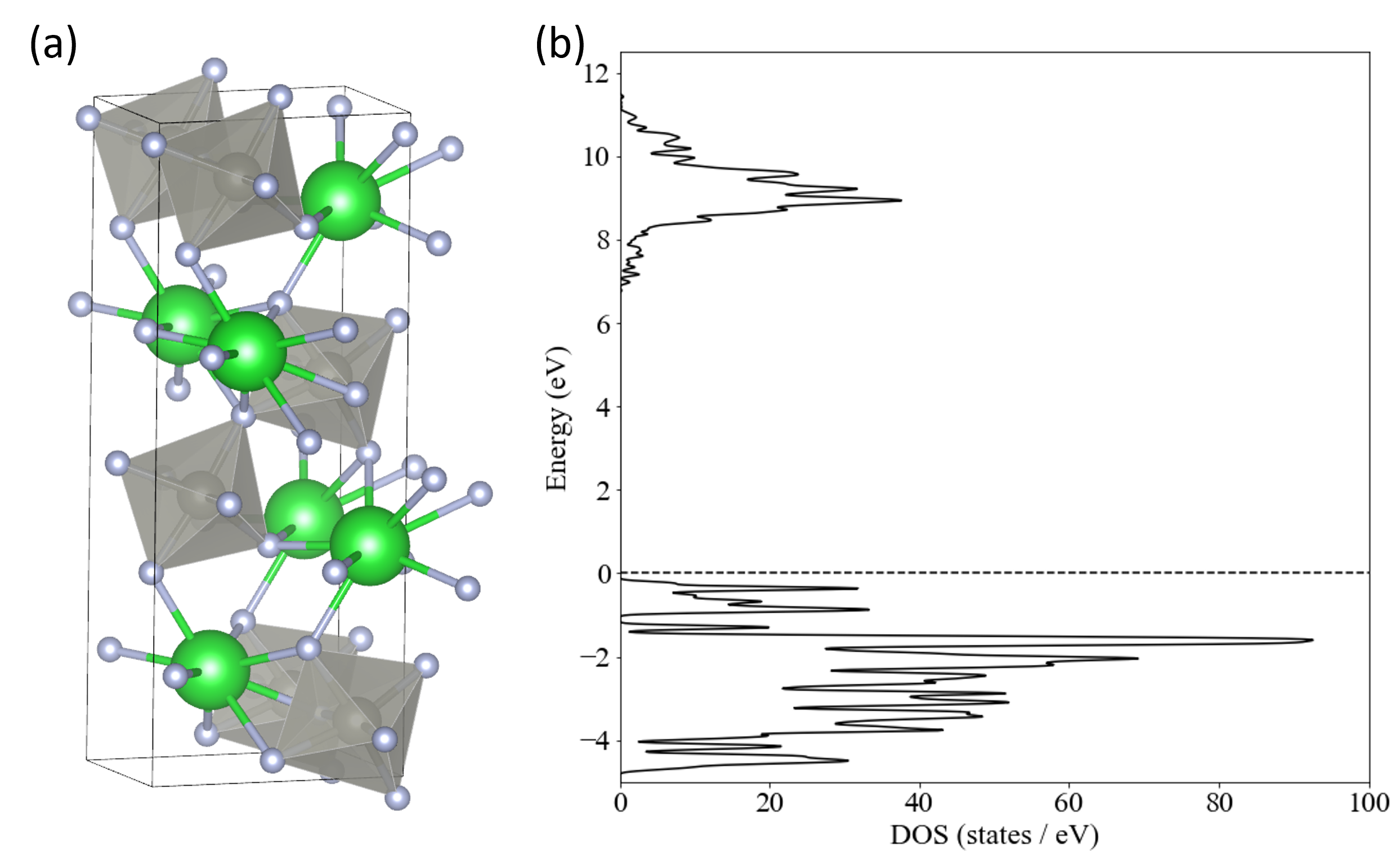}
    \caption{(a) Unit cell of \ce{BaZnF4}. Ba is shown in green, Zn in dark grey, and F in light grey. (b) Total density of states (DOS) of \ce{BaZnF4}.}
    \label{fig:BaZnF4 bulk}
\end{figure}

The same set of calculations was performed for \ce{BaZnF4}, which has the same structure as \ce{BaMgF4}.
The optimized structure and DOS are shown in Figure \ref{fig:BaZnF4 bulk}.
The computed band gap is 7.00 eV, an underestimate compared to the experimental (absorption edge) value of 8.55 eV.\cite{RN541}
Since measurements of the band gap suggest that it is in fact slightly higher than the nuclear transition energy, we analyzed the energy of Th defects in the material.

The three defect cells equivalent to those studied in \ce{BaMgF4} were constructed; Th replaces Ba in all cases, and the charge is compensated by either \ce{V^{$\prime\prime$}_{Ba}},  \ce{V^{$\prime\prime$}_{Zn}}, or \ce{2F^{$\prime$}_{i}}.
The calculated properties of these defects are summarized in Table \ref{tab:BaZnF4 defects}.
\begin{table}[htb]
\begin{tabular}{c|c|c}
Defect   & $\Delta E_{f}$ (eV) & $E_d$ (eV) \\
\hline
\ce{Th^{..}_{Ba} + V^{$\prime\prime$}_{Zn}} & 0.68     & 4.44          \\
\ce{Th^{..}_{Ba} + V^{$\prime\prime$}_{Ba}} & 1.05     & 5.01         \\
\ce{Th^{..}_{Ba} + 2F^{$\prime$}_{i}}       & 1.28     & 5.36         
\end{tabular}
\caption{Calculated formation energies $\Delta E_{f}$ and Fermi energy to first defect state energies $E_d$ of thorium defects in \ce{BaZnF4}.}
\label{tab:BaZnF4 defects}
\end{table}
The formation energies are all lower than their equivalents in \ce{BaMgF4}, suggesting that \ce{BaZnF4} may host a higher concentration of thorium.
The relative energies of the defects follow the same trend, with a \ce{Zn^{2+}} vacancy being the most likely charge compensating defect, followed by a \ce{Ba^{2+}} vacancy, and \ce{F-} interstitials.
Thorium PDOS plots for the defect cells are shown in Figures S4-S6.
They show that the electronic properties of thorium in Th:\ce{BaZnF4} are much the same as in Th:\ce{BaMgF4}.
In the cation vacancy cells the Th $5f$ orbitals form narrow peaks at 4-6 eV, though the majority of the thorium-containing states in the valence band are more than 2 eV below the Fermi level so the electronic transitions involved in internal conversion may be higher energy.
Nevertheless, given the clear gap between the thorium $5f$ band and the conduction band, and the fact that the experimental band gap is only just above the nuclear transition energy, we suspect that internal conversion through the Th $5f$ orbitals could be possible in Th:\ce{BaZnF4}.

The difference in defect stats of \ce{BaMgF4} and \ce{BaZnF4} results from the valence orbitals of \ce{Mg^{2+}} and \ce{Zn^{2+}}.
Zinc lies at the end of the first transition series, so \ce{Zn^{2+}} has a $3d^{10}4s^0$ electronic configuration.
The $3d$ orbitals are chemically inert, in that there are no generally accepted reports of \ce{Zn^{3+}} compounds, but the PDOS plots in Figures S7 show that the top of the valence band is principally composed of Zn $3d$ orbitals, somewhat hybridized with F $2p$~\cite{RN560,RN561}.
This contrasts with the valence band in \ce{BaMgF4} which is entirely F $2p$ because the highest occupied orbital in \ce{Mg^{2+}} is the core-like $2p$.
In both cases the conduction band is predominantly Ba $5d$, though the band gap of \ce{BaZnF4} may be somewhat reduced by the low-lying $4s$ orbital of \ce{Zn^{2+}}.

As a result, the two materials have fundamentally different excited electronic states.
In Th:\ce{BaMgF4} the first excited state is an electron transfer from \ce{F-} to \ce{Th^{4+}}, similar to what is expected in other fluorides like \ce{LiSrAlF6} and \ce{CaF2}.
However, in Th:\ce{BaZnF4} the hole created by electronic excitation will be centered on zinc, likely with significant hybridization with fluorine.
The different excited state characters could affect wavefunction overlap and lead to different nuclear quenching behavior in the two materials.

\subsection{DFT methods}
DFT calculations on were performed with VASP\cite{RN12}, version 6.4.2, using the PAW\cite{RN14} method with a plane-wave cutoff of 500 eV and a spin-restricted formalism.
The PBE\cite{RN13} functional was used for all structural optimizations, and the modified Becke-Johnson (MBJ)\cite{RN489,RN490} functional was used for electronic properties.
Structural optimizations were performed using a $\Gamma$-centered $k$-meshes with point spacings of 8-2-6 for bulk \ce{BaMgF4} and \ce{BaZnF4}, and 2-2-2 for the defect supercells, all equivalent to a $k$-point spacing of 0.03 \AA{}$^{-1}$.
Structural optimizations of other binary fluorides also used a $k$-point spacing of 0.03 \AA{}$^{-1}$.
MBJ calculations on the bulk unit cells used 8-2-6 $k$-point grids, and MBJ calculations on the defect supercells used 3-3-3 $k$-point grids.
$k$-point grids were generated with VASPKIT.\cite{VASPKIT}

\section{Optical Contact Bonding}
Figure \ref{fig:bmf_concept_art} outlines a composite optical crystal for optical doubling and excitation of \ce{^{229}Th}.
In this system the incident laser first reaches pure a nonlinear material to generate 148 nm photons.
These photons then travel into a \thor{}-doped crystal to excite the \ce{^{229}Th}.

Clearly, this requires that the interface between the two halves of the crystal be transparent to photons.
Assuming no adhesives are used to bond the crystals, they must be held together by van der Waals interactions~\cite{RN616}, a process known as 
`optical contact bonding'.

For the bonding to work the crystal surfaces must be exceptionally flat and clean.
Surface preparation typically involves extensive polishing, and the actual contacting procedure may be done under vacuum.
Additional processing steps can improve the quality of the contact.
High pressure can be applied to try to overcome surface roughness.
Annealing, where energy is supplied to the system for the atoms close to the contact to rearrange into a more stable configuration, can be done with heat, electricity, or light.
A more recent development is `chemical activation' of the surface.
The basic principle is to disturb the surface significantly at the atomic level, creating `dangling bonds' that will easily form covalent bonds between the surfaces when they are joined.
This can be done by surface chemical reactions, such as hydroxy-catalysis bonding in silicate glasses, or physical processes, such as ion beam or plasma etching~\cite{RN617,RN618}.

Several factors indicate that bonding of a \ce{BaMgF4}/Th:\ce{BaMgF4} composite is likely to be successful.
It involves bonding doped and undoped crystals of the same material so the lattice constants and coefficients of thermal expansion are near-perfectly matched.
In a relevant study, a composite system of pure and Nb-doped YAG was successfully produced by optical contacting including surface activation by argon atom beam etching.\cite{RN618}
Fluorides can be used in optical contact bonding, including \ce{YLiF4} and \ce{LuLiF4}, as can non-linear optical materials~\cite{onyxopticsMATERIALCOMBINATIONS}.


\section{Example Clock}
As an example of a potential architecture for solid-state clock, we consider the example of \bmf{} probed with a single $\sim$296.8 nm laser, as shown in Fig.~\ref{fig:bmf_concept_art} (a). 
The clock transition frequency is primarily determined by the isomer shift, the crystal electric field gradient, and the local magnetic field. 
The isomer shift, also known as the chemical shift in Mössbauer spectroscopy, is primarily associated with the interaction between the nuclear charge distribution and the electronic environment surrounding the nucleus. 
As a result, the apparent nuclear transition frequency varies across different hosts and doping sites. 
We estimate the isomer shifts by summing over the  partial density of states integrated over the valence band (IPDOS)  for various partial waves ($s$, $p$, $d$, and $f$) weighted by the isomer shifts for the valence atomic states of  \thor{}$^{3+}$ ion. 
This procedure yields isomer shifts relative to the apparent nuclear clock frequency for the \thor{}$^{4+}$ closed-shell ion. 
\thor{}$^{4+}$  serves as a convenient reference as its electronic cloud
remains largely unperturbed by the crystal fields as supported by our density of state calculations. 
We find that the isomer shifts range from $\delta_{\textrm{iso}} = -254$~MHz to $-258$~MHz for BaMgF$_4$ materials and over a much wider range, $\delta_{\textrm{iso}} = -224$~MHz to $-268$~MHz for BaZnF$_4$ due to a much larger IPDOS variations across  BaZnF$_4$ substitutions. 
These values are for the lowest formation energy defects compiled in Tables~\ref{tab:BaMgF4 defects} and~\ref{tab:BaZnF4 defects}.
For comparison, we find the isomer shift of \thor{}:\lisaf{} and \thor:\ce{CaF2} as $-248$~MHz and  $-243$~MHz, respectively.

\begin{figure}
    \centering
    \includegraphics[width=\linewidth]{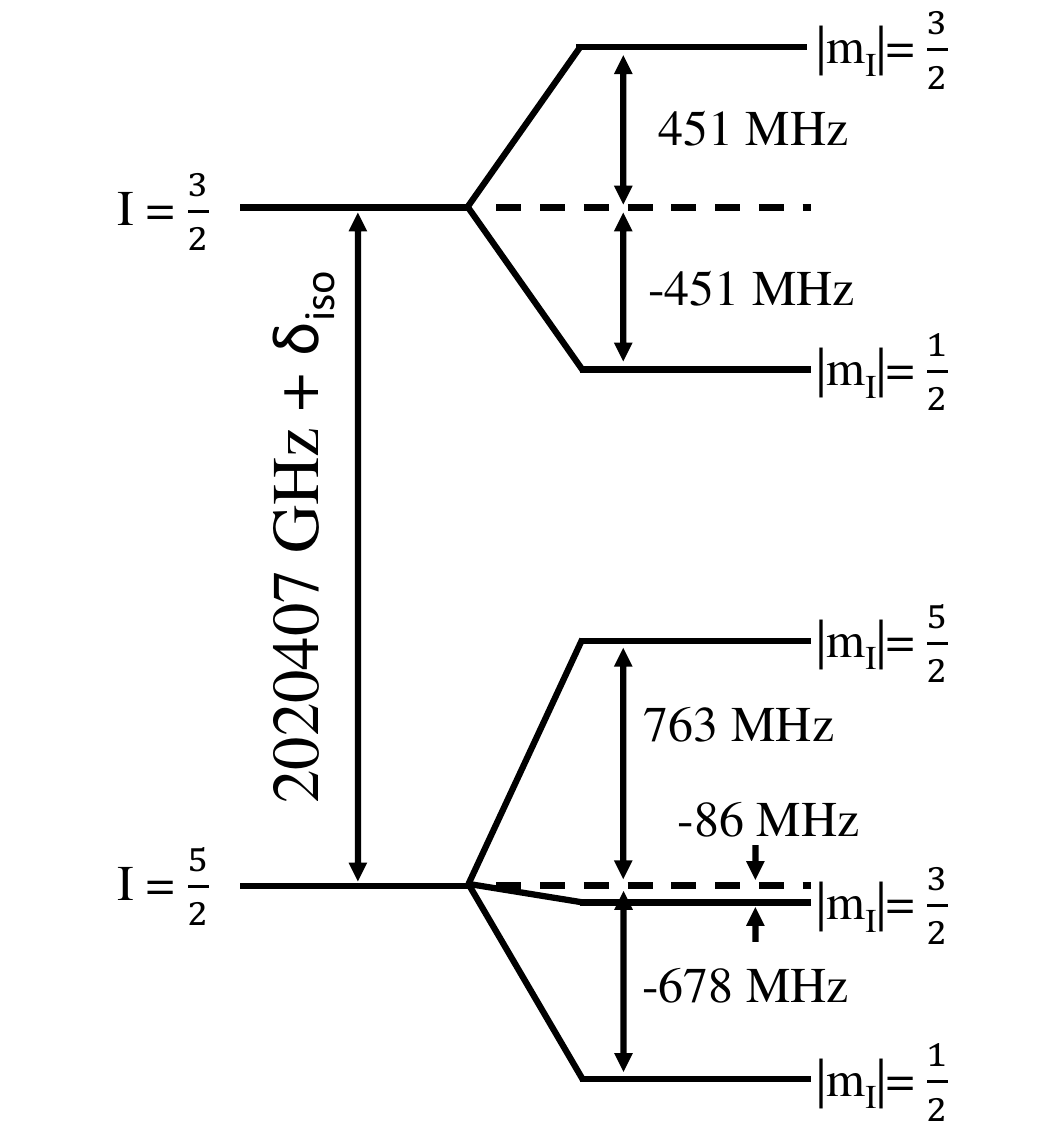}
    \caption{Splitting of the $|m_I|$ levels of the nuclear ground and isomeric state in the electric field gradient of \bmf.}
    \label{fig:efg_splitting_zeeman_splitting}
\end{figure}

In addition to the isomer shift, the main effects of the crystalline host on the \thor{}$^{4+}$ nucleus are the coupling of the \thor{} nuclear electric quadrupole (EQ) moment to the crystal electric field gradients (EFG) $\{V_{xx},V_{yy},V_{zz}\}$ and the coupling of the \thor{} nuclear magnetic moment to the magnetic field created by the other atoms in the crystal~\cite{Rellergert2010}. 
\thor{}:\bmf{} is an ionic crystal with no unpaired electrons and therefore the magnetic field experienced by a \thor{} atom comes from the other nuclear magnetic moments in the crystal. 
Thus, the nuclear energy levels are described by the Hamiltonian:
\begin{equation}
    \hat{H}= -\mu_\alpha \vec{I}\cdot\vec{B} + \frac{eQ_\alpha V_{zz}}{4I(2I-1)}\left[3\hat{I}_z^2-\hat{I}^2 +\eta(\hat{I}_x^2-\hat{I}_y^2)\right], \nonumber
\end{equation}
where $\hat{I}$ is the total nuclear spin operator, $\hat{ I}_{x,y,z}$ are the component of the nuclear spin operators, $\eta = |V_{xx}-V_{yy}|/V_{zz}$ is the EFG asymmetry parameter with the choice $|V_{zz}| > |V_{xx}| > |V_{yy}|$, and $\alpha = \{g,e\}$ denotes the ground and excited nuclear states, respectively, which is parameterized by $Q_g = 3.149(4)$~eb~\cite{Bemis1988}, $\mu_g = 0.360(7)\mu_N$~\cite{Gerstenkorn1974, Campbell2011}, $Q_e = 1.77(2)$~eb~\cite{Thielking2018, Yamaguchi2024}, and $\mu_e = -0.378(8)\mu_N$~\cite{Thielking2018, Yamaguchi2024}, where $\mu_N$ is the nuclear magneton.
The DFT calculations described above predict $V_{zz}$~=~398~V/\AA{}$^2$ and $\eta = 0.585$, leading to the splitting of the nuclear levels as shown in Fig.~\ref{fig:efg_splitting_zeeman_splitting}.
Using the DFT predicted positions of the F atoms in the \bmf{} crystal, we find that fluctuations in the magnetic field at the position of the \thor{} nucleus leads to an inhomogenous broadening of the $\ket{I = 5/2, m_I = \pm1/2} \leftrightarrow \ket{3/2,\pm1/2}$ transition of $\approx$700~Hz. 
Based on the measurement in Ref.~\cite{Zhang2024}, we predict this transition occurs at approximately $2020407371(2)$~MHz. 

To produce light at this frequency, we assume efficient frequency doubling requires quasi-phase matching in the \bmf{} crystal.
Given that the largest nonlinear doubling coefficient is $d_{32} \sim 0.06$ pm/V~\cite{Buchter2001}, we take the input polarization along the crystal $b$-axis. 
From the Sellmeier coefficients provided in~\cite{Buchter2001}, the indices of refraction are estimated as 1.645 at 148.4~nm, and 1.472 at 296.8~nm. 
This yields a poling period of $\Lambda = 2\pi/\Delta k \approx 858$ nm.
Sub-micron poling has been demonstrated in lithium niobate thin films~\cite{slautin2021}, however, to our knowledge it has not been extended to bulk crystals. 
As such, we consider the doubling efficiency based on both a $\Lambda$- and $3 \Lambda$-poled crystal.

The intensity $I_2$ of second-harmonic radiation generated in a doubling crystal with $m \Lambda$-poling is given by~\cite{Boyd_NLO_2020}
\begin{equation}
    I_2(2 \omega) = \frac{2 d_m^2 (2\omega)^2 I_1^2}{n_1^2 n_2 \epsilon_0 c^3} L^2,
\end{equation} where 
\begin{equation}
    d_m =  \frac{2 d_{32}}{\pi m}.
\end{equation} 
If we assume $I_1=200$ mW/mm$^2$ of 296.8 nm light into a $L=5$ mm \bmf{} crystal one will obtain $5.5 \times 10^{-7}$ mW/mm$^2$ of 148.4 nm light for $\Lambda$-poling and $6.1 \times 10^{-8}$ mW/mm$^2$ for 3$\Lambda$-poling. For the remaining analysis the input light is assumed to have a 1 kHz linewidth.

Assuming that the generated light enters a 1~cm \bmf{} region doped with \thor{} at a doping density of $10^{16}$ cm$^{-3}$, the ultimate limit on the clock stability is given by~\cite{MartinBoydThesis2007}
\begin{equation}
    \sigma = \frac{1}{2 \pi Q S}\sqrt{\frac{T_e + T_c}{\tau}},
\end{equation} where $Q=f_0/\Delta f$ is the transition quality factor, $S$ is the signal-to-noise ratio (SNR), $T_e$ is the excitation time, $T_c$ is the fluorescence collection time, and $\tau$ is the averaging time. The SNR is given by
    $S = \frac{N_{d}}{\sqrt{N_d + b \times T_c}}$,
where $b=200$ cps based on backgrounds measured in \thor{}:\lisaf crystals, and $N_d$ is the number of detected photons given by 
    $N_d = \eta N_e \left( 1 - e^{-\Gamma T_c}\right)$,
where $N_e$ the total number of \thor{} nuclei excited, $\eta$ = 0.01 is the assumed system detection efficiency, and $\Gamma$ is the transition decay rate. 
Optimizing $\sigma$ with respect to ($T_e$, $T_c$), we find the ideal combinations as (570~s, 570~s) for $\Lambda$ and (580~s, 550~s) for $3\Lambda$. From the assumed parameters we obtain a projected clock performance of $5 \times 10^{-16}/\sqrt{\tau}$ and $2 \times 10^{-15}/\sqrt{\tau}$ for $\Lambda$ and $3\Lambda$-poling, respectively.


\section{Conclusion}
In order to advance the development of a solid-state nuclear clock, we analyzed the viability of \bmf{} and \bzf{} for generating VUV radiation at the \thor{} isomer wavelength of 148.6 nm. In addition, we propose to directly dope \thor{} into a VUV doubling crystal as a potential ``vacuum-free" VUV laser clock or to join the VUV doubling and \thor-doped crystal together through optical contacting. In line with the former proposal, we have analyzed the potential shifts and broadenings that the crystalline environments could introduce to the isomeric transition, and have estimated the clock performance of an all-\bmf{} nuclear clock.

\begin{acknowledgments}
This work was supported by NSF awards PHYS-2013011, PHY-2207546, and PHYS-2412982, and ARO award W911NF-11-1-0369.
ERH acknowledges institutional support by the NSF QLCI Award OMA-2016245.
This work used Bridges-2 at Pittsburgh Supercomputing Center through allocation PHY230110 from the Advanced Cyberinfrastructure Coordination Ecosystem: Services \& Support (ACCESS) program, which is supported by National Science Foundation grants \#2138259, \#2138286, \#2138307, \#2137603, and \#2138296.
\end{acknowledgments}

\section*{Data Availability Statement}
The data that support the findings of this study are available within the article [and its supplementary material].

\bibliography{Th229-apd,ThoriumSearch,ref,HM_refs}

\end{document}